**Generative Artificial Intelligence in Higher Education: Evidence from an Analysis of Institutional Policies and Guidelines**

Nora McDonald, Aditya Johri, Areej Ali, Aayushi Hingle

**Abstract**

The release of ChatGPT in November 2022 prompted a massive uptake of generative artificial intelligence (GenAI) across higher education institutions (HEIs). HEIs scrambled to respond to its use, especially by students, looking first to regulate it and then arguing for its productive integration within teaching and learning. In the year since the release, HEIs have increasingly provided policies and guidelines to direct GenAI. In this paper we examined documents produced by 116 US universities categorized as high research activity or R1 institutions to comprehensively understand GenAI related advice and guidance given to institutional stakeholders. Through an extensive analysis, we found the majority of universities (N=73, 63%) encourage the use of GenAI and many provide detailed guidance for its use in the classroom (N=48, 41%). More than half of all institutions provided sample syllabi (N=65, 56%) and half (N=58, 50%) provided sample GenAI curriculum and activities that would help instructors integrate and leverage GenAI in their classroom. Notably, most guidance for activities focused on writing, whereas code and STEM-related activities were mentioned half the time and vaguely even when they were (N=58, 50%). Finally, more than one half of institutions talked about the ethics of GenAI on a range of topics broadly, including Diversity, Equity and Inclusion (DEI) (N=60, 52%). Overall, based on our findings we caution that guidance for faculty can become burdensome as extensive revision of pedagogical approaches is often recommended in the policies.

1. **INTRODUCTION**

The introduction of new technology into higher education is often a catalyst for significant changes in teaching and learning. Although the overall evidence on the role technology, especially computing technologies, plays in improving the quality of teaching and learning is debatable [19] each major computing innovation is met with fervor within higher education. This is true of technologies adapted for education as well as those designed solely for teaching and learning [5]. In recent years, the use of Artificial Intelligence (AI) driven applications and systems have become a major topic of both education research and practice. For researchers, the intersection of AI with already growing fields of educational data mining and learning analytics has created new opportunities to both understand, predict, and intervene in educational and learning outcomes [30, 36].

For educators, the use of applications for writing and coding, among others, have made AI an important consideration. Not surprisingly, researchers have started to investigate the use of AI across a range of educational topics and applications such as simulation-based learning [20], curriculum design [46], poetry writing [29] and storytelling [34]. One recent application that has found special resonance within higher education is ChatGPT, a large language model (LLM) driven application that can be used through a chat-based interface. ChatGPT is one form of language-based Generative AI (GenAI) in an increasingly crowded category. Due to its intuitive ease of use the release of ChatGPT [40] quickly ushered in an era of AI use across higher education institutions (HEIs) forcing faculty and administrators to respond to an increase in academic misconduct, especially the use of the application by students for assignments requiring text based responses.

Like most technology applications before this, ChatGPT has evoked mixed reactions. The initial reaction from institutions ranged from a spectrum of calls to ban it, on one end, and to embrace it fully, on the other. The Internet was overrun with advice on how to use ChatGPT and how to catch those using it. There has also been a wholesale embrace of GenAI by some for its possibilities to do everything from polishing prose to writing emails and articles (Kalla & Smith, 2023) [28]–and in this vein, overcome language barriers, structural barriers and inequalities [49]. Overall, responses to ChatGPT and other language-based GenAI in our culture and in our classrooms, in particular, have ranged quite a bit. This has resulted in a slew of guidelines for teachers and students on how to use LLM based applications.

In addition to teaching and learning, the introduction of ChatGPT has also introduced an occasion to reexamine existing research practices, especially as it pertains to formulation of research problems, data analysis, and writing. Researchers anticipate that GenAI applications like ChatGPT (and its successors) that are based on large language models (LLMs) will improve coding, save time with reading, writing, and administrative tasks, and otherwise augment research and help to generate ideas [45]. In a survey of postdocs, GenAI was considered a big boon to writing and coding [35] and analysis of social media discussions suggest that the general sentiment towards GenAI is positive and that we only need clear policies about use [17].

One reliable and important source of information regarding the use of GenAI for faculty and students are policies put forward by different institutions. Across HEIs, a range of responses could be seen with some institutions releasing brief guidance and others elaborative policies and instructions. Most recently, an investigation similar to this one was done of the GenAI policies of the world's 50 top-ranking HEIs [33] and found that guidance generally covers "academic integrity, advice on assessment design and communicating with students." Moorhouse et al also find that HEIs have largely come to embrace GenAI, a position which the authors themselves advocate. The goal of this study was to explore what policies universities in the US with high research activity (R1 institutions) have adopted around GenAI, what guidance they are giving faculty instructors, and to identify themes across institutions. Data used in this study revealed most universities supply a range of guidance on the uses of GenAI in the classroom, from prohibition to wholesale endorsement. Despite this flexibility, universities trend towards embracing GenAI to the point of endorsing a revision of pedagogical approaches, with no regard for the long term pedagogical implications, time investment of instructors and students, changes to notions of ownership, and with less than half discussing the privacy implications and ethics of the use of GenAI.

## 2.    RELATED LITERATURE

Although the release of ChatGPT and other GenAI applications for education is quite recent, there is already significant scholarly literature on the topic. In this section we discuss some of the issues brought forth by scholars and educators to help us frame our analysis and discussion.

## 2.1. Resistance to ChatGPT

Although attitudes and norms seem to be constantly in flux, there are voices in education that are staunchly opposed to LLM-based GenAI applications like ChatGPT [15, 48]. Some have been sounding the alarms about the entrance of GenAI and its impact on teaching and learning [13, 38]. Others have taken umbrage at the suggestion that ChatGPT or related applications be used by instructors for lesson planning or to create any other content [25], or that students would be given a tool to plagiarize or counterfeit their knowledge that was even more undetectable. The release of ChatGPT was quickly followed by the introduction of ZeroGPT, a detection tool that since its introduction has been both a leader of the "arms race" [10], and by no means a silver bullet [21, 47].

During the initial period after the release of ChatGPT, academics and institutions reacted with urgency to combat ChatGPT similar to their usual response to plagiarism [41]. There were even glimmers of hope that, for instance, because ChatGPT "hallucinates," [3] and makes up citations and sometimes loses its way in longer writing, it would be easy to detect its use. Since then, a number of GenAI detection tools have come on the market. But general consensus is that (at least for now) these tools still remain unreliable [16]. Perhaps, partly as a consequence, there seems to be somewhat of a shift away from this combative approach to GenAI [33]. While articles and formal guidance now abound about the way ChatGPT detectors are not very good, very recent work suggests that this battle may not be over [39]. Indeed, even while schools may have temporarily (or permanently) abandoned the effort to detect AI, the race continues to produce better detection tools that substantially outperform popular text classifying tools like those produced by the makers of ChatGPT, Open AI, and ZeroGPT [39]. And of course, there are ever more creative new modes of detection (e.g., prompt engineering) that are emerging.

## 2.2 Cautious Embrace of ChatGPT

By contrast, there are those who embrace ChatGPT in the classroom, enthusiastic about its possibilities for transforming pedagogy, lesson plans, and efficiency. Some have seen the introduction of ChatGPT as a needed nudge to already outmoded practices–for example, the end of the term English essay in place of more relaxed and impromptu sharing of reflections and journal writing [24]. Some scholarly writing has been dedicated to showing how GenAI can improve teaching and help instructors attain, before unreachable goals [27]. For example, the United Nations Education, Scientific and Cultural Organization (UNESCO) suggests it can be used as a "personal tutor," "study buddy," and "socratic opponent" among other things [14]. Many nevertheless caution that its benefits can only be obtained when balanced with good teaching (e.g., [18, 23]) and many questions still remain [37].

Harvard's metaLab has released an AI guide that provides background and tools for educators to use in the classroom [2]. In addition to suggesting that educators be clear about their AI policy–whether students can use it, where, and how to cite it–they make suggestions for how to incorporate it in the classroom, including incorporating a "critical component" into assignments with AI so that students can assess how well or not and its influence on students own thinking. They also suggest reflecting on the implications of AI for "equity, democracy, education, or information quality" [2]. The amount of meta-thinking around the insertion of GenAI tools is something we reflect on in the discussion. Needless to say, those using it must do a lot of additional work.

Mollick and Mollick argue that ChatGPT can improve "transfer" (application of knowledge to other contexts), "evaluation" (to deepen knowledge) and "breaking the illusion of explanatory depth" (the impression one has that they possess more knowledge than they do) [32]. Mollick and Mollick offer exercises that are reproduced and adapted in some of the documents in our analysis. In addition to deepening learning through class assignments and exercises, some have suggested using GenAI for tailored learning and tutoring.

In their systematic review of research on GenAI in educational domains including engineering education, Bahroun et al. found work on the use of GenAI in learning environments. For instance, papers suggested support for automation of curriculum design and enhanced learning outcomes with use of chatbots like ChatGPT in exams and peer feedback [8]. They also highlight the pitfalls of plagiarism, bias, copyright, and inclusivity, discussed later in this section.

Yet, we also know that in some educational settings, instructors are struggling to understand GenAI's capabilities and feasibility in their classroom [9, 42]. Tan and Subramonyam found that to help teachers and instructors better understand the vast uses of GenAIs like ChatGPT it is important to allow them to discover the teaching and learning possibilities that GenAIs offer. GenAIs like ChatGPT impact directly the learning outcomes of students because when there is uncertainty surrounding how they can be used this leads to over monitoring, which can result in serious consequences for student learning outcomes [42].

While certainly GenAI has the potential to save time in the classroom and also reduce some of the workload needed to prepare to teach and administer assessments [9] it's still not clear if this is beneficial in the long term to the engagement and knowledge building of instructors. Even among those embracing it, there are reservations and discussion of requirements for ongoing discussion to ensure transparent use of GenAI [13].

Some have also importantly pointed out that GenAI cannot address structural problems in teaching and academic research (e.g., pressures to produce research and publish) [39]. GenAI will not, for instance, repair our overburdened system or structural biases or other deficiencies in education.

**2.3 Privacy Concerns**

Another area of concern that has been discussed in the literature is its implications for privacy. GenAI raises concerns about confidentiality and privacy both in teaching and research. In teaching, one primary concern would be that entering students assignments into GenAI might be a violation of the Family Education Rights and Privacy Act (FERPA) which is a law in the US that limits access to educational information and student records. Under FERPA, school employees may not divulge information about student grades or behavior with anyone. For instance, they cannot post student grades on a public bulletin board. While universities have contracts with platforms like Blackboard and Canvas, where they indeed post student grades, those platforms are designed to support FERPA and also work with individual school policies [44]. It is not clear under what circumstances the use of GenAI violates FERPA and/or school privacy policies.

Another issue is that encouraging students to use GenAI means they will be giving up personal information to a third-party. This may be no different then allowing or encouraging use of collaborative tools like those in Google Drive. At the same time, it's possible that some more highly sensitive data could be easily coupled with student email and be part of a data breach or shared with third-parties that track and target students with ever more personal ads. Through the lens of surveillance capitalism [50], GenAI seems optimized for privacy incursion.

## 2.4 Intellectual Property Concerns

There is also the murkier issue of whether it is ethical to have students produce content with GenAI, even with attribution to GenAI applications, because the content is itself taken from (trained on) other sources. This is why well known authors are suing ChatGPT for copyright infringement [4]. There is arguably something problematic in requiring students to cite sources but only asking them to cite ChatGPT broadly, suggesting that it is only authenticity, not intellectual property, that is at issue. Also at issue is the fact that when students enter essays in ChatGPT for the purposes of improving prose (or for any other reason) their own intellectual property is being sent out into the GenAI ether.

## 2.5 Diversity, Equity and Inclusion (DEI)

Like many other internet-based technologies, there is much optimism for GenAI blazing a path for equity–for instance, providing writing support for those whose first language is not English, or other skills and learning that are only available to those who can afford education. At the other end of the spectrum are worries that it could widen gaps by supporting already privileged students, particularly those who can afford updated versions [17] and also reinforcing existing biases about students' abilities [43].

## 2.6 Research Considerations Related to ChatGPT

While it's not clear how widespread use of GenAI is in academic research papers–and indeed some have admitted to using it without disclosing that in publications– some academics report having embraced tools like ChatGPT to brainstorm and clarify ideas and writing [16]. Among those many scientific publications that have responded with policies, some have banned it while others require transparency [16]. Some have cited benefits like "equity" arguing that GenAI will support non-English speaking researchers to overcome barriers to publishing their work [45]. The issues are, of course, complex and even while GenAI presents potential boons to academic writing and research efficiency, ethical concerns about intellectual integrity, as well as its more subtle erosion of "authenticity and credibility" are unsettled [22]. At issue is not just that GenAI can produce entirely fake publications [31] but its use as "ghost author" [26].

**2.7 GenAI and the Future of the Workforce**

As universities understand the various uses and ways to integrate GenAIs into curriculum and how best they would serve student learning and course development, it is also important to remember GenAIs have permanently impacted education and will continue to do so [7]. As universities understand how to implement policy and procedure for the use of GenAIs it is imperative to consider how to prepare students for professional spaces where GenAIs will be a part of their work [7]. It is important that universities are teaching students about the uses of GenAI and the ethical implications this will have on their work and the world as a whole [12]. Notably, much of the concern about the introduction of GenAI into education and academia has been focused on writing and research, but not Engineering or STEM. Rarely do we see conversations about how GenAI is going to transform math or programming learning– for better or worse.

**2.8 Research Questions**

Based on our review of prior work, the following research questions guided our data selection and analysis:

RQ1: What guidance are higher education institutions (HEIs) providing to their constituents about the use of GenAI?

RQ2: What is the overall sentiment on the use of GenAI articulated by HEIs (e.g., do they encourage or discourage use) and how is it manifested in actual guidelines?

RQ3: What ethical and privacy considerations, if any, are represented in the guidelines?

**3. RESEARCH STUDY**

Given our objective to understand how HEIs are approaching GenAI, for our data we decided to collect related policy documents and guidelines that have been released and made public on the internet. This approach is similar to the one followed by [11]. In their study, they were interested in understanding policies that govern data use within HEIs given the proliferation of data-driven applications. To achieve their goal, they conducted a policy discourse analysis of 151 university policy statements related to student information privacy and the responsible use of student data from 78 public and private post-secondary institutions in the United States (US). They identified prominent discourses related to privacy solutions, institutional responsibility and student agency, and discussed the consequences of these discourses. They limited their study to US institutions as norms, laws, and regulations vary across countries and policy making is often unique within a given context. In our study, we adopt a similar approach both in terms of data collection and analysis. We further limited the institutions to R1 to achieve higher consistency in data interpretation.

### 3.1 Data Collection

The data collection for this project started first with identifying a sample set of institutions from where to gather information about their generative artificial intelligence policies. The Carnegie Classification is a framework for classifying colleges and universities in the United States. In the first stage of data collection, the Carnegie list of institutions that are classified at R1 institutions was used. R1 institutions are universities with the highest level of research activity. This list contained 131 institutions. After a list of institutions was established, the researchers searched through each university's website looking for their policy (or mention of) generative "artificial intelligence" ("AI" or "ChatGPT") use in the classroom. If there was no information on the university's website about their policy on GenAI or mention of GenAI, then researchers also conducted a Google search to confirm that there was no missed information. Researchers only used what was publicly available and did not obtain documents only available through institutions resources (e.g., links to Google Docs). The data was collected from October 9, 2023 to November 26, 2023. This search was not fully exhaustive and could exclude policies published or made publicly available after the collection period, even if the publication date is a date prior or during the collection time period. There were ultimately 14 institutions that had no policy or mention of GenAI use in the classroom at the time that we searched, resulting in a total of 116 institutions that were included in our analysis. Among those 116 institutions with information about a GenAI policy or mention of GenAI in the classroom, we downloaded all pages that specifically dealt with GenAI for a total of 141 documents. These data were then open coded by each institution.

### 3.2 Open Coding

All four researchers reviewed a random subset of the data (N=20 institutions) and initially discussed codes and developed a codebook. After reviewing another random sample of N=10, this codebook was refined. Two researchers then applied this codebook to another random set of 10 institutions, establishing near perfect agreement to gain inter-rater reliability (IRR). Given the open-ended, qualitative nature of the analysis we did not use formal reliability measures. The two coders then coded the remaining dataset separately using an initial codebook (see Table 1). The initial analysis identified five main areas (see Figure 1). Additional validation was done during a subsequent coding round in which subcodes were applied to initial codes. These subcodes are shown in Table 1 and are elaborated on in our findings.

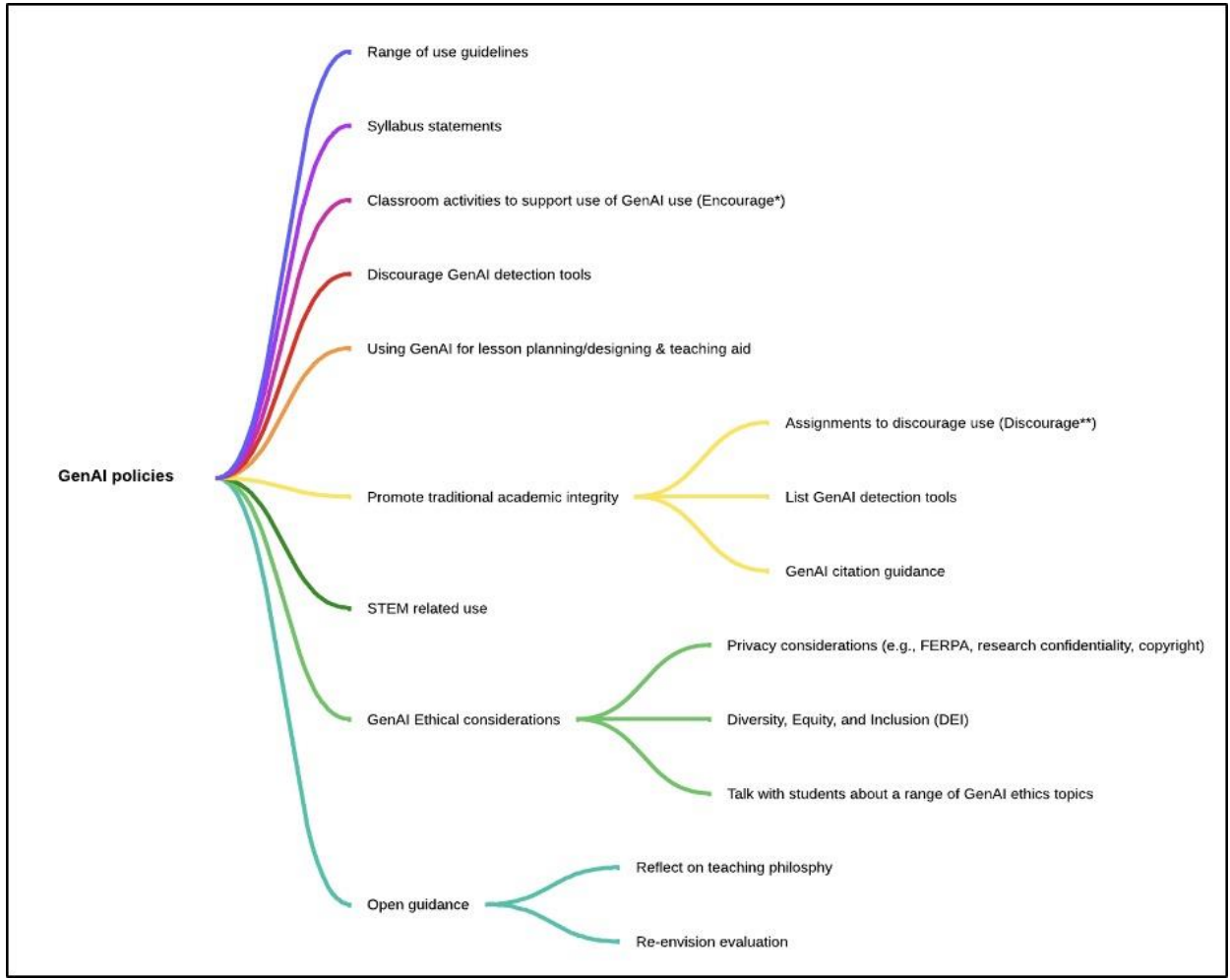

GenAI policy codes applied to the dataset. See Table 1 for description of subcodes.
*We ultimately classified those who provided classroom activities that supported use as "encourage" use along with those that did not discourage use but provided 4 or more of these codes. **Those with this code were classified as "discourage" use. See Figure 2 for more details.

**Figure 1**

## 4. FINDINGS

### 4.1 Overview

More than half of all institutions in our analysis provided suggested sample syllabi for faculty (N=65, 56%). More than half stipulated syllabus statements for a range of use (N=64, 55%), often in three categories such as "embrace," "limit," or "prohibit" (N=43, 37%) vs those that provided more narrow or broader use guidance (e.g., "permissible" and "prohibited" or "no use," "some use," "unlimited use," and "required use") (N=21, 18%).

While the majority of institutions offered a range of options for faculty, there were those that leaned in more to embracing GenAI. Half (N=58, 50%) provided sample GenAI curriculum and activities that would help instructors integrate and leverage GenAI in their classroom, and more than two in five (N=51, 44%) discouraged use of GenAI detection tools. Notably, one third of institutions even provided guidance for faculty on how to use GenAI for lesson planning (N=34, 30%).

By contrast, more than half (N=63, 54%) provided guidance on designing assignments that discouraged use of GenAI by students and more than one in five (N=27, 23%) provided guidance on using GenAI detection tools.

Half suggested that teachers reflect on their approach to teaching and evaluating students (N=58, 50%) leaving it open as to how they approached it for their classroom. There is some overlap between institutions that offer both types of guidance on how to embrace and discourage use. In some instances, this reflects institutions' desire to provide support for faculty based on their intentions to use (or not use) GenAI. It also reflects a general wariness that strategies to discourage use, detection tools in particular, will be outpaced by rapid innovation.

Both stances, encouraging and discouraging use, imply rethinking of the classroom, however–and much work for instructors and students. What is indeed clear is that many institutions seem determined to provide instructors with guidance about how to flip their classroom and otherwise rethink their teaching and evaluation strategies.

Slightly more than one half of institutions talked about the ethics of GenAI on a range of topics, including Diversity, Equity and Inclusion (DEI) (N=60, 52%), privacy with GenAI–particularly concerns about entering sensitive data or third-party data sharing among GenAI platforms (N=66, 57%)–and the need to have discussions with students about the ethics of using GenAI in the classroom (N=61, 53%).

Notably, most guidance for activities focused on writing, whereas code and STEM-related activities were mentioned half the time and vaguely even when they were (N=58, 50%).

In the next section we describe each of these themes in more depth.

| Codes / themes | N= | Subcodes / subthemes |
|---|---|---|
| **Range of use guidelines** | 64 | • Provide three GenAI use guidelines for instructors (e.g., "embrace", "limit", "prohibit")<br>• Provide narrower or broader set of GenAI use guidelines for instructors (e.g. "No use," "Some use," "Unlimited use," |
| | | "Required use" or "permissible" and "prohibited") |
| **Syllabus statements** | 65 | • Provide language for instructors to include in their syllabi according to the various use guidelines |
| **Classroom activities** | 58 | Exploring strengths and weaknesses of GenAI (e.g., through evaluation of veracity, fact checking, writing). Using GenAI for support to deepen discussions, debate with, brainstorm, draft, etc. Teaching GenAI skills (e.g., prompt refinement or "prompt engineering") "Turing tests" to assess whether something is GenAI or not |
| **Discourage GenAI detection tools** | 51 | Discourage use, with or without including GenAI detection resources (see below "List of GenAI detection tools") |
| **Lesson planning** | 34 | Creating and getting feedback about activities (e.g., designing activities, games, getting feedback about how students will respond) Creating lesson plans, lectures, teaching tools Creating quizzes, test questions, projects Customize learning (Providing custom feedback, curriculum, and learning tools for students) Others: responding to student email, creating accessible content, translation, professional development |
| **Assignments to discourage use** | 63 | Use of course materials and core concepts or other contextualization (e.g., personal details) Flip-classroom and process-based evaluation Thinking-centric approach where learning is made visible (e.g., reflective prompts) Scaffolding assignments (small steps, edits, refinements, that build toward larger project) Leveraging GenAI content (critiquing, comparing text, identifying bias, incorrect answers) Higher order thinking Test evaluations with GenAI to refine and make more difficult Compensate for GenAI with low stakes assignments that help with retrieval Collect writing samples |
| **List GenAI detection tools** | 27 | 18 out of 23 explicitly mention tools – most often Turnitin (N=16), ZeroGPT (N=8), Hugging Face (N=3) and a few others less frequently mentioned |
| **Privacy** | 69 | Caution with sensitive or private information, copyright Caution about legal implications (FERPA, HIPAA, etc) Concerns about students privacy/personal information for |
| | | account<br>Suggest providing opt-out for students worried about privacy<br>Review/link to ChatGPT privacy policies |
| **Talk with students** | 61 | Suggestion that instructors be clear about their policies and talk to students about the ethics of using GenAI including:<br>● The learning process and benefits of critical thinking<br>● Academic integrity and plagiarism<br>● Bias and inaccuracy of GenAI output<br>● Data privacy concerns<br>● Dependency/reliance on GenAI<br>● Intellectual-based skill atrophy (e.g., complex problem-solving)<br>● Implications of Human-AI collaboration<br>● Loss of learning opportunities/quality |
| **Open guidance** | 70 | ● Encourage instructors to reflect on what GenAI means for their teaching and re-envision evaluation |
| **DEI** | 60 | ● Bias in GenAI output can harm underrepresented or underprivileged students through the usage of microaggressions<br>● Accessibility and equity: make accommodations for internet access, payment and consideration of other disabilities and barriers to accessing GenAI<br>● Assist with language for students students for whom English is not their first language |
| **STEM related use** | 58 | ● Mention use in:<br>○ Computer Science<br>○ Math<br>○ Natural sciences, healthcare, engineering, other |
| **Include citation reference format** | 44 | ● APA, MLA, and some other like Chicago, IEEE<br>● Custom format (e.g.,"Chat-GPT-3. (YYYY, MM, DD of query) 'Text of your query.'") |

A table showing codes applied to universities policies on GenAI (far left column) with Ns for those codes (middle column), and related subcodes listed as bullets (far right column).

**Table 1**

## 4.2 Embracing GenAI

Many institutions are embracing GenAI by providing instructors with sample activities for the classroom, ideas for lesson planning, and the admonition that GenAI detection tools are faulty and unreliable. These themes and their subthemes (shown in Table 1) are detailed in this section.

### 4.2.1 GenAI curriculum and activities

Institutions that are generally embracing GenAI advise instructors to explore its strengths and weaknesses by having students evaluate its veracity ("Creating text for students to fact check, practicing disciplinary research skills" (UChicago)) and writing quality. Others might suggest asking GenAI to generate statements that were analyzed for "inaccurate, misleading, incomplete, and/or unethical information" (Yale University). A common suggestion is that instructors build in the use of GenAI to support and deepen group discussion and help students individually brainstorm (e.g., "Brainstorming: AI can be especially useful before students start drafting. For example, AI can provide background information or generate a list of research questions" (UC Davis)), help with "brainstorming or drafting" (Columbia University), or debate ideas, acting as a kind of "socratic" tutor.

A smaller subset of institutions also emphasize building students' skills with GenAI through "prompt engineering" where students learn to refine their prompts (or queries). UCLA instructs teachers to "use prompting logic used by students to generate information and then provide a different prompt to help guide revision. Showcasing that small changes can lead to major differences in output!" (UCLA)

Frequently, GenAI is being suggested as a companion to generate critical ideas on specific topics that are then treated as content to be critiqued by the class. Take for example, this suggestion by University of Oregon:

> "Students can also practice analyzing and synthesizing texts using ChatGPT. For example, the instructor can task ChatGPT with generating a comparative essay that analyzes the gendered norms in "The Story of an Hour," by Kate Chopin, and "The Yellow Wallpaper," by Charlotte Perkins Gilman. Once the essay is generated, students can analyze its synthesis of the two stories and identify opportunities for improvement."

Another version of this offered by University of Southern California was to "[a]sk students to complete a written assignment, then use AI to generate a version of the same assignment. Instruct students to compare the two and reflect on their work." Other schools embrace the notion that instructors should "re-envision" writing as a process of editing and refinement (e.g., University of Central Florida). Others suggest treating GenAI like a "study buddy" (University of Pittsburgh) or turn to it for "creative inspiration" (University of Arkansas) or to "reduce anxiety" when producing a first draft (Columbia University).

### 4.2.2 Discourage use of GenAI detection tools

Many institutions that mention AI detection tools caution instructors that they are not reliable. Even the those applications like Turnitin that instructors have long relied on for plagiarism detection are cast in doubt: "Although companies such as Turnitin are beginning to offer AI detection services, none have been established as accurate" (Carnegie Mellon). By contrast, some, like Case Western, couch their guidance in the acknowledgement that these tools have never been reliable saying that "scholars have long debated the merits and limitations of plagiarism detection software (e.g., Turnitin), and these conversations continue with the advent of new AI-detection tools." They go on to recommend that faculty be cautious and thoughtful about how they include detection software in their courses.

The vast majority of universities that provide resources for AI detection make explicit mention of tools. Turnitin is mentioned most often (N=18), followed by GPTZero (N=8), Hugging Face (N=4) and a few others less frequently mentioned (e.g., openai-detector, Gltr.io).

### 4.2.3 Lesson Planning

Encouraging the use of GenAI for instructor lesson planning was somewhat less common but still discussed by a sizable number of institutions (N=34, 29%). The most common guidance was to use GenAI to design and get feedback about classroom activities and assistance with lesson planning content creation (e.g., creating slides, lecture material, etc.). The University of Central Florida instructs faculty to "[m]ake the AI your teaching assistant. When preparing a course, ask the AI to explain why commonly-wrong answers are incorrect. Then, use the Canvas feedback options on quiz/homework questions to paste the AI output for each question. Teach sentence diagramming and parts of speech." A number of institutions suggest that instructors use GenAI to learn how students might respond to their assignments (e.g., "AI Prompts for Teaching is a compendium of prompts that instructors can use to discern how GenAI would respond to their course assignments, materials, etc." (University of Mexico)).

Many among this group of institutions also suggested instructors use it for lecture design (e.g., "Ask Chat GPT to synthesize text from large documents. For example, enter a 3500 word paper as a prompt, and ask ChatGPT to create an 18-slide PowerPoint presentation, with headings and bullet points, making a persuasive case for action" (UCLA)) and to design quizzes, projects, exams, as well as rubric. Yet a smaller group described its utility in providing tailored learning and feedback for students (e.g., "Personalized Recommendations and Content Curation-AI algorithms can analyze student preferences, past performance, and learning patterns to recommend relevant resources, supplementary materials, or additional learning opportunities" (University of Hawaii)).

What is remarkable is the amount of effort that both teachers and students are directed to give to incorporating GenAI into their classroom activities and assignments. Students are encouraged and/or required to not only use GenAI in various ways but also to describe the nature of their use of it, providing details about their queries and, above all, cite it–more than a third (N=44, 38%) provide formal citation guidelines, which are most often references to the APA style guidelines.

### 4.3 Discourage GenAI

While a number of institutions provide guidance for curriculum involving GenAI, a good number (N=31, 27%) also provide curriculum to discourage use.

### 4.3.1 Assignments to discourage use of GenAI

Among those institutions providing guidance for instructors in how to limit or prevent use of GenAI in their classroom, most mention providing assignments that require engagement with course material or core concepts as well as requesting other contextualizing details (e.g., personal input or reflections from students.)

> "For example, ask students to apply concepts, solve problems, or analyze case studies in ways that integrate class discussions, lectures, lived experience, and specific course readings" (University of Oregon).

The same number of institutions suggested that instructors wholeheartedly embrace the "flip classroom" approach–in essence, fundamentally change their approach to teaching. Flip classroom suggestions included process-based evaluations like oral exams, debates, discussions. This overlaps somewhat with the "thinking-centric" approach where learning is made more visible with reflective prompts, like the suggestion by University of Salt Lake City to "[u]tilize reflection and metacognition to have students respond to assignments that indicate that they have done more than follow prompts."

Many among this subset of those discouraging GenAI suggest instructors scaffold assignments with smaller steps that build toward larger projects (e.g., "By adopting iterative assessment practices, instructors can engage students in ongoing dialogue, offer guidance, and evaluate progress. This approach discourages students from relying solely on AI-generated content as it becomes evident that growth and improvement depend on active learning rather than quick fixes through AI tools." (George Mason University)). Indeed, Wayne State tells instructors that "AI (so far) is not very good at scaffolding work from one assignment to another, so any time you can build writing assignments that build on prior work, it's more difficult to rely on AI to produce meaningful content. Topic proposals, intro paragraphs, drafts and revisions, any kind of scaffolding is difficult to fake with AI."

Somewhat less frequently discussed but still noteworthy was emphasis on appealing to "higher order" thinking with "[q]uestions that require logical reasoning: Questions that involve deductions, inferences, or abstract concepts may be hard for the model to answer." (University of Lincoln).

In contrast to those institutions that encourage GenAI by suggesting instructors use GenAI to refine their questions, some in this group of institutions focused on testing evaluations with GenAI to make the questions more difficult.

Notably, not all institutions providing guidance to discourage use are those that are recommending detection tools. For instance, University of Central Florida suggests that perhaps discouraging use through other means is the only recourse, and cautions that preventing students from using GenAI is a futile exercise because "the technology is improving rapidly, and automated detection methods are already unreliable (at UCF, in fact, the office of Student Conduct and Academic Integrity will not pursue administrative cases against students where the only evidence is from AI detectors)." Indeed, AI, the University of Central Florida argues, "is simply a fact of life in modern society, and its use will only become more widespread."

## 4.4 Privacy

About three in five institutions (N=69, 60%) caution instructors about privacy concerns with GenAI. Almost every institution that talked about privacy concerns with GenAI advised instructors to exercise caution about sharing personal or sensitive data with GenAI. Fewer (less than one in five) talked about making their students aware of the risks of how their data is being used (e.g., "Ensure that any personal or sensitive information shared during AI interactions complies with privacy regulations. Implement necessary measures to protect data and inform students about how their information is being used" (University of Houston)) and refrain from entering personal data.

Indeed, admonitions about privacy are quite frequently vague and not student-focused:

> "ChatGPT is a for-profit tool, actively gathering data from users who input information. … As such, if instructors choose to use ChatGPT for their teaching, they assume responsibility for reviewing and vetting concerns with accessibility, privacy, and security. Instructors should remain open to giving students alternative options for completing an assignment if ChatGPT is inaccessible to them in any way." (Berkeley)

> "Material that you submit may then become part of the program's database—using the software also contributes to its development … Even software platforms that don't use your content to train the product DO collect your information and often install cookies to track your other activity. While you may choose to make this bargain to try out AI tools yourself, it can be more problematic to require students to sign up for tools that will track and use their private data." (Yale University)

A common theme is to place the burden elsewhere and suggest that students "may be uncomfortable," as opposed to signaling an ethical stance: "...some students may be uncomfortable using or creating accounts with generative AI applications." (Ohio State University)

Some institutions were more explicit about data flows involving sensitive user information as well as concern for intellectual property (both their own and the data used) and how to instruct students, like the University of Pittsburgh which gave instructors prompts for considering intellectual property rights:

> "Learn who can use or own the data that AI tools receive as input or produce as output. Large language models store input to use as training data. Explain to students that AI tools store and use their data and avoid inputting student-generated content into AI without their consent. Consider and discuss potential copyright concerns with students. (University of Pittsburgh)

A number of institutions suggested that instructors give students the option to "opt out" of using GenAI, citing privacy concerns and cost (e.g., "Opt-out options: If creating assignments that require students to use AI tools, be aware of costs and privacy concerns. In recognition of privacy and equity concerns related to the use of AI tools, students should have the option to opt out of using AI." (University of Maryland)). Brandeis suggested that if instructors "choose to use chatGPT in your course, one approach to avoiding these privacy concerns is to ask students to use the program via anonymous email accounts and to allow students to opt-out." Johns Hoipkins similarly suggested "Having students use a pseudonym and/or generic email address." The University of Oregon suggested that instructors remind students of the risk and provide an opt out as they do with other external vendor tools:

> "We therefore strongly recommend that instructors who ask or encourage students to use any AI system remind students that they should avoid providing any personal or other sensitive data to AI prompts. We also advise that instructors consider making AI use voluntary or, if AI use is part of a required course assignment or activity, include an opt-out alternative for students who do not want to create an account with an AI system or interact with them."

When it comes to using GenAI for grading, some institutions suggested that instructors read ChatGPT's privacy policy as well as consider the legal implications of entering student data (e.g., "Regardless of data collection and retention policies, USC researchers, staff, and faculty should be particularly cautious to avoid sharing any student information (which could be a FERPA violation), proprietary data, or other controlled/regulated information." (University of Southern California)). It was indeed surprising given that some institutions are encouraging instructors to use GenAI for grading and curriculum that more emphasis was not placed on prohibitions on entering student data–only 18% mention the legal implications.

> "The main concern with sharing student work relates to students' privacy rights. FERPA protection begins after an instructor accepts an assignment for assessment and grading." (Carnegie Mellon)

> "Submitting student information to an AI system could constitute a violation of FERPA unless the system is approved for such use through university agreements" (University of Miami).

### 4.5 Diversity, Equity, and Inclusiveness (DEI)

The institutional guidance around GenAI and DEI that comes up most frequently is the potential for biased output with GenAI and the need for accommodations for students who are underprivileged and face barriers to accessing the internet, paid subscriptions for GenAI, as well as consideration for other disabilities that might present barriers to use. Overcoming language barriers and consideration of larger societal harms with GenAI (or AI in general) with respect to structural bias, labor practices for training are sometimes also mentioned.

When discussing bias in GenAI output, some speak of "harmful" and "biased" content that students could encounter, with some framing it specifically as a result of training on publicly available data that "perpetuate biased, discriminatory, or inaccurate information" (University of Pittsburgh) or "on a large body of writing created by humans with human biases the responses generated could reflect and further enforce those same biases." (University of Southern Florida).

For accommodations, there is somewhat of a range of idiosyncratic guidance centering around financial hurdles and visual disabilities. University of Maryland specifically talks about financial burdens as well as vision impairments saying: "Some students may not be comfortable signing up for accounts in these tools, or may experience barriers to access if they use tools such as screen readers" and goes on to suggest that instructors have "a plan for a student who needs to opt out of using a tool." University of Nevada, stipulates that AI can help students with accessible and customized learning that supports their pace and learning needs, or provide remote learning opportunities but notes that people don't always have equal access to infrastructures that would support it, or as Carnegie Mellon points out, "consider whether or not they are digitally accessible to all learners."

Notably, University of Nebraska says that GenAI can be used to support those with vision impairments: "Generative AI is being developed to take existing images and enhance clarity, adjust color, and translate the content in many other ways to enhance accessibility. This can allow students to take the content you create and tailor it to their specific needs. It is still essential for instructors to take steps to create content that is as accessible as possible, but the ability for individuals to get more out of the content we create brings a bit of excitement to the experience." They also point out the bias inherent in AI technology which is trained on data that underrepresents women and people of color, citing the use of AI in the medical field as an example.

Those who pointed out the benefits of GenAI for non-native English speakers do sometimes mention a paradox: while GenAI can help with grammar and language, those same students who rely on GenAI for that kind of help may more often be subject to accusations of plagiarism. Other echoes of this have surfaced in our reading of University blogs where instructors point out that, in a world where it is assumed to be used, minority students (even those who don't use GenAI) are now more likely to be accused of plagiarism with the tool because of structural bias.

In all of these DEI policies it is not always clear whether they are discussing GenAI or AI technology more broadly.

## 4.6 STEM-related use

Half of the institutions mention use of GenAI in STEM related courses (N=58, 50%) with most mentioning computer science (N=56, 48%) and fewer discussing math or natural sciences. Engineering is only mentioned by N=7 institutions.

Even when STEM courses are mentioned, however, it is quite frequently superficial with institutions talking about the role of GenAI in helping with "coding," often alongside "writing" or, less often, in a long list of fields which are associated with non-ChatGPT tools (e.g., "However, many other AI generative tools automate content generation across all different fields, music, voice, imagery, computer code, etc." (University of Utah)). Indeed, there is a sense that coding and related STEM applications are not seen as a major threat or boon. Take for instance, Yale University, when they remind instructors that GenAI will have an "impact on learning" that "applies across all disciplines" but says that "STEM problem sets that require explanations also depend on students' generating language to learn more deeply."

### 4.7 Research

University policy rarely mentions use of GenAI in research. Less than ten institutions mention it in the context of privacy (data not shown) but often with the caveat that they do not have a good sense of the privacy implications. For example, the University of Oregon cautions faculty that they do not have contracts with GenAI platforms and are not sure if they are in compliance with their academic, business or research but instruct faculty to "avoid unintended release of intellectual property, copyrighted materials or trade secrets." Similarly, the University of Central Florida says that faculty should be "cautious, if not outright paranoid, about privacy, legality, ethics, and many related concerns, when thinking about exposing your primary research to any AI platform."

The University of Southern California advises "researchers, staff, and faculty" to "be particularly cautious to avoid sharing … proprietary data, or other controlled/regulated information." Washington University in St Louis tells faculty that they simply should "not enter confidential or protected data or information, including non-public research data, into publicly available or vendor-enabled AI tools."

### 4.8 Where do institutions fall along the spectrum of embracing (or not) GenAI?

In our review of documents, we labeled institutions as "discouraging" use if they seemed to suggest that GenAI was not welcome in the classroom with curriculum designed for those purposes or aggressively emphasized the dangers (legal and intellectual) of using it. More than a quarter (N=31, 27%) fell into this category, though some while providing little guidance at all (N=9, 7%). (See Figure 2). We find that the majority of universities (N=73, 63%) encourage use and supply ample guidance for uses of GenAI in the classroom (N=48, 41%). Others we labeled as encouraging use do not necessarily provide sample curriculum, but do give extensive guidance around GenAI, including syllabus statements, that instructors reflect on their use of GenAI, ethics guidance, etc. (N=25, 22%). (We defined "encourage" use as those that do not discourage use and either supply ample guidance on its use in the classroom or provide ample guidance in the other categories in Table 1.) A similar number provided no sample curriculum or extensive guidance in any other categories but did not seem to discourage use, and we labeled them as in between (N=12, 10%).

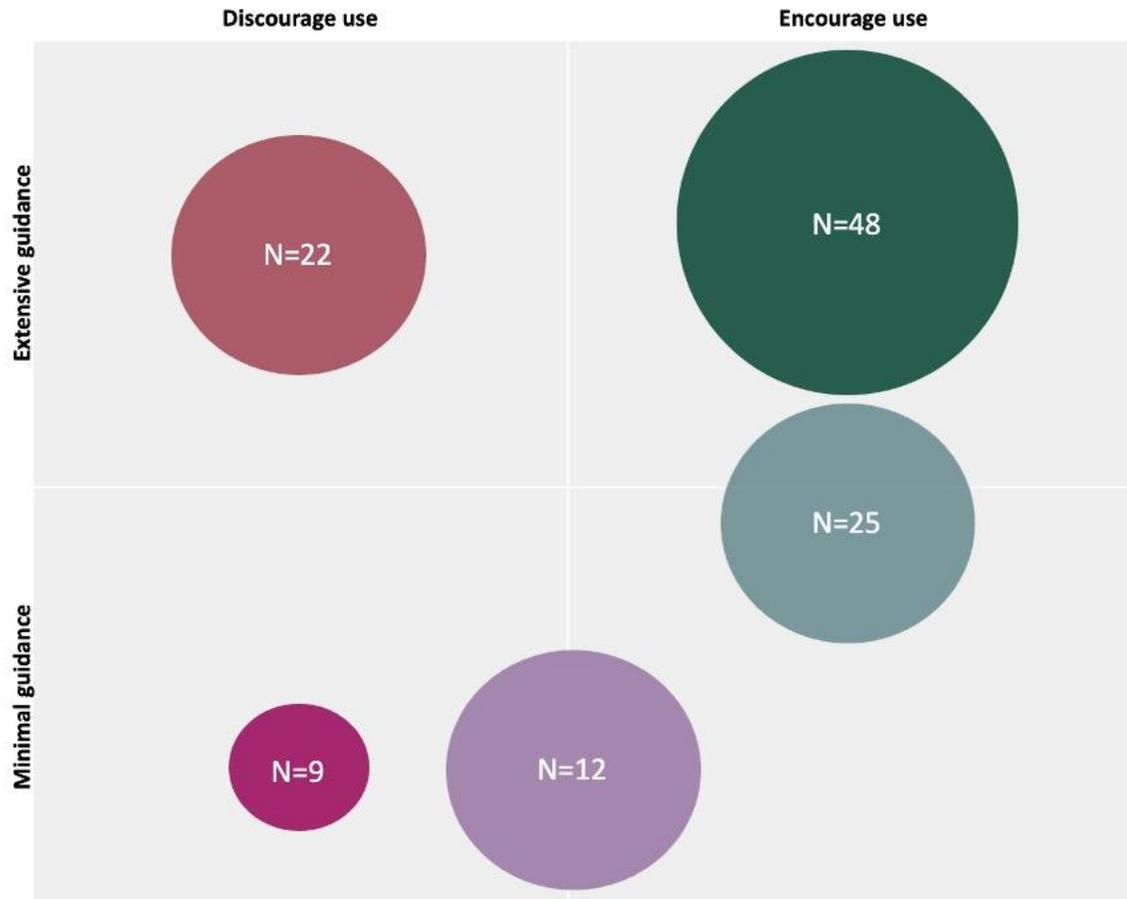

The 4 Quadrants show "discourage use" and "encourage use" (left and right) and "extensive guidance" and "minimal guidance" (top and bottom). University policies in each of these 4 categories (and one in between) are represented with bubbles and Ns.

**Figure 2**

## 5. DISCUSSION

Based on our analysis, a sizable number of institutions embrace GenAI, even while still providing instructors with some strategies to limit the use of GenAI. Some acknowledge their embrace is largely pragmatic (e.g., "simply a fact of life in modern society," no detection tool will ever be satisfactory, etc.), others position GenAI as an opportunity to rethink teaching and learning.

Regardless of where institutions fall on the spectrum, the result is many policies that model a classroom in which GenAI is the protagonist. GenAI is positioned as a tool for group work, for engagement, to deepen discussion and debate, as well as for providing students with custom tutoring and feedback. A good number of institutions suggested GenAI for lesson planning and a few even suggested that instructors use it to grade, to answer email, and customize resources for individual student learning. Overall, our analysis shows that GenAI is being heralded by many institutions as a tool that can solve many problems or be used to increase productivity across teaching and learning. But we are skeptical that this will actually happen–and that, in fact, what has been introduced is more work (and perhaps, less trust) for all.

A common theme across institutions that embrace GenAI is an attempt to confront larger structural issues that impact teaching, especially plagiarism. Plagiarism and intellectual dishonesty have confronted higher education for a long time and many solutions to the "problem" have been advanced. Increasing use of surveillance during testing is one common solution. But the heavy use and growing sophistication of plagiarism and surveillance testing software has nevertheless told a different story–of a preference for the status quo, particularly amidst a pandemic.

A different route that has been taken is to redesign how teaching takes place and "flipped" classrooms was one recent trend in this category. The idea was that by working on assignments during class there would be more accountability as evaluations will take place during instructional time and under instructional supervision with no opportunity or incentive for dishonesty [1, 6].

Perhaps ironically, GenAI seems to have pushed institutions (or given them no other choice, or simply the justification they needed) to embrace *GenAI flip* models. The problem is that, as we have reported, GenAI is not necessarily the impetus for adopting in-classroom evaluations, reading and reflections, or oral examination (as may have been envisioned with flip model) but is, in fact, *the centerpiece* of learning– where the skills that are being taught are primarily to *query, prompt-engineer, and criticize a chatbot* and where the outputs are frequently not students own work. If, for example, according to much of the guidance given, scholarly output now involves meta commentary on how GenAI has been used, the burden of evaluators shifts from authentication and learning assessment to documentation and critique of prompt engineering.

While evolving the classroom is ostensibly a good thing, as noted, there are reasons to be wary. First, many institutions whose guidance encourages the use of GenAI express little to no concern for ethics and privacy associated with using it–some even present guidance that is in direct contradiction, suggesting for example that instructors use GenAI to grade but also to exercise caution when entering sensitive data. FERPA violations, intellectual property, and student privacy are the concern of a subset and this concern often seems an afterthought.

Second, in an effort to discourage use, many institutions seem to have, paradoxically, overly embraced GenAI, figuring it into every aspect of curriculum. The prescription for GenAI as a "buddy" or a "tutor" aside, its presence in the classroom as a *central character and artifact* is troubling. GenAI is not a student. So many institutions suggested activities that positioned GenAI as an object to be critiqued, compared, fact-checked, etc. This is, in fact ,what we should ask of ourselves, absent GenAI. Institutions seem so concerned about GenAI's entrance that they confuse acknowledging it with experimenting with it in the classroom. In our analysis, there is no reference to research backed evidence for the guidance that is given.

Third, the thrust towards normalization of GenAI through constant use runs the risk of making its presence indiscernible to students and to teachers who are being instructed to embrace GenAI and also monitor its use–heaping on tremendous burdens of time and ethical challenges. There must be considerable cognitive overhead (for both students and teachers) that accompanies turning in assignments with GenAI prompts and other process data. We might also worry about the intellectual discouragement that accompanies doing an assignment and then asking GenAI to replicate it for the purposes of comparison.

Fourth, and relatedly, it's just not clear what the long term impact on intellectual growth and pedagogy will be. Indeed, there is the larger issue of depending on GenAI for critical analysis, idea generation, and brainstorming. Some institutions appear to acknowledge the risks, though miss the mark, by reminding instructors to use activities that reinforce retrieval–implying that this is a skill that will undoubtedly atrophy as we strengthen our GenAI muscles.

## 6. CONCLUSIONS, LIMITATIONS, AND FUTURE WORK

We conducted a survey of GenAI policies to see what guidance institutions are giving their faculty. We found themes related to communication and curriculum guidance, lesson planning, privacy and ethics. We characterized institutions based on the degree to which they encouraged or discouraged use and how much guidance they offered. One important insight we offer is that many institutions are fervently embracing GenAI, and in a way that presents potential burdens for faculty and students and without much regard for ethical concerns. We also argue that encouraging the use of GenAI the way many institutions do may not produce the learning outcomes desired. Looking forward, the introduction of GenAI can be seen as a catalyst for changing assessment and evaluation practices that are more ecologically valid and grounded in principles of fairness, justice, and ethics, but these positive outcomes require, perhaps, a more thoughtful consideration of its role in the classroom.

While we have tried to capture GenAI policies at a critical moment in time, it still represents one moment in time. Moreover, we could not capture all of the guidance that instructors are exposed to at their institutions and among their colleagues and networks. Future work will explore similar higher education contexts outside the US.

**ACKNOWLEDGEMENTS**

This was partially funded by NSF grant #2319137.